\documentclass[aps,showpacs,nofootinbib]{revtex4}

\usepackage{latexsym}
\usepackage{epsfig}
\usepackage{amssymb}

\newcommand{\lp}{\left(}
\newcommand{\rp}{\right)}
\newcommand{\lb}{\left[}
\newcommand{\rb}{\right]}

\newcommand{\ba}{\begin{eqnarray}}
\newcommand{\ea}{\end{eqnarray}}
\newcommand{\be}{\begin{equation}}
\newcommand{\ee}{\end{equation}}

\newcommand{\al}{\alpha}

\newcommand{\ka}{\kappa}

\newcommand{\Ga}{\Gamma}

\newcommand{\R}{\mathcal{R}}

\newcommand{\OA}{\Omega_A}

\newcommand{\F}{\OA+\phi}
\newcommand{\Fp}{\lp\OA+\phi\rp}

\begin{document}

\title{Cosmology of hybrid metric-Palatini $f(X)$-gravity}

\author{Salvatore Capozziello$^1$}\email{capozzie@na.infn.it}
\author{Tiberiu Harko$^2$}\email{harko@hkucc.hku.hk}
\author{Tomi S. Koivisto$^{3}$}\email{tomi.koivisto@fys.uio.no}
\author{Francisco S.N.~Lobo$^{4}$}\email{flobo@cii.fc.ul.pt}
\author{Gonzalo J. Olmo$^{5}$}\email{gonzalo.olmo@csic.es}

\affiliation{$^1$Dipartimento di Scienze Fisiche, Universit\`{a} di Napoli "Federico II", Napoli, Italy
INFN Sez. di Napoli, Compl. Univ. di Monte S. Angelo, Edificio G, Via Cinthia, I-80126, Napoli, Italy}
\affiliation{$^2$Department of Physics and Center for Theoretical
and Computational Physics, The University of Hong Kong, Pok Fu Lam Road, Hong Kong}
\affiliation{$^{3}$  Institute of Theoretical Astrophysics, University of
  Oslo, P.O.\ Box 1029 Blindern, N-0315 Oslo, Norway}
\affiliation{$^4$Centro de Astronomia e Astrof\'{\i}sica da Universidade de Lisboa, Campo Grande, Ed. C8 1749-016 Lisboa, Portugal}
\affiliation{$^5$Departamento de F\'{i}sica Te\'{o}rica and IFIC, Centro Mixto Universidad de
Valencia - CSIC. Universidad de Valencia, Burjassot-46100, Valencia, Spain}

\date{\today}

\begin{abstract}

A new class of modified theories of gravity, consisting of the superposition 
of the metric Einstein-Hilbert Lagrangian with an $f(\R)$ term constructed 
\`{a} la Palatini was proposed recently. The dynamically equivalent 
scalar-tensor representation of the model was also formulated, and it was 
shown that even if the scalar field is very light, the theory passes the Solar 
System observational constraints. Therefore the model predicts the existence 
of a long-range scalar field, modifying the cosmological and galactic 
dynamics. An explicit model that passes the local tests and leads to cosmic 
acceleration was also obtained. In the present work, it is shown that the 
theory can be also formulated in terms of the quantity $X\equiv \kappa^2 T+R$, 
where $T$ and $R$ are the traces of the stress-energy and Ricci tensors, 
respectively. The variable $X$ represents the deviation with respect to the 
field equation trace of general relativity. The cosmological applications of 
this hybrid metric-Palatini gravitational theory are also explored, and 
cosmological solutions coming from the scalar-tensor representation of 
$f(X)$-gravity are presented. Criteria to obtain cosmic acceleration are 
discussed and the field equations are analyzed as a dynamical system. Several 
classes of dynamical cosmological solutions,  depending on the functional form 
of the effective scalar field potential, describing both accelerating and 
decelerating Universes are explicitly obtained. Furthermore, the cosmological 
perturbation equations are derived and applied to uncover the nature of the 
propagating scalar degree of freedom and the signatures these models predict 
in the large-scale structure.

\end{abstract}


\maketitle

\section{Introduction}

Recently, a novel approach to modified theories of gravity was presented,
consisting of adding to the Einstein-Hilbert Lagrangian an $f(\R)$
term constructed \`{a} la Palatini \cite{Harko:2011nh}. Using the respective
dynamically equivalent scalar-tensor representation, it was shown that
the theory can pass the Solar System observational constraints even
if the scalar field is very light. This implies the existence of a
long-range scalar field, which is able to modify the cosmological
and galactic dynamics, but leaves the Solar System unaffected. An explicit model that besides passing the local tests leads to cosmic acceleration was also presented.
The main motivations to this approach lies on the fact that metric and geodesic structures of gravity could play important roles in dynamics as clarified by Ehlers-Pirani-Schild (see \cite{eps} and reference therein for a comprehensive discussion on this problem). Furthermore, the good experimental results of general relativity, at local scales and in the weak field approximation, should be retained by any theory of gravity aimed to explain large scale and cosmological dynamics as prescribed by for example the {\it chameleon mechanism} \cite{camel} or by the {\it disformal screening mechanism} \cite{Koivisto:2012za}.

In this context, there has been a considerable interest in the modifications 
of the geometric part of the Einstein's field equations, motivated mainly
by the observation of the late-time accelerated expansion of the
Universe \cite{expansion}. In particular, gravitational actions
consisting of more general combinations of curvature invariants
than the pure Einstein-Hilbert term have been investigated
extensively \cite{fRgravity,revnoi}. Einstein himself was more satisfied
with the geometric part of his equations, and has been quoted to
say that while the left hand side is carved of marble, the right
hand side is made by straw. However, in generalized gravity
theories, the problem of coupling matter to gravity is often
reduced to the question of in which frame matter resides with
respect to gravity. The matter Lagrangian, and the corresponding
stress-energy tensor, are defined in the usual way, but the metric
that matter couples to can be related to the gravitational metric by a
conformal, or, more generally, by a disformal transformation.
This is a fundamental issue in relation to observations, essentially based on matter dynamics. Specifically, if one is adopting a Jordan frame matter is minimally coupled to geometry. On the other hand, in the so called Einstein frame, matter can result non-minimally coupled to geometry and the gravitational coupling can depend on time and distances \cite{cqg}. At least in the case of non-universal coupling, this fact can pose the fundamental problems of what the {\it true physical frame} is and where theory is consistent with data (see \cite{libro}  for a comprehensive discussion). However, apart from some non-conservation terms in the continuity
equations, the general structure of the theory is retained.

New light on this old problem is shed within the recently introduced framework 
denoted C-theories \cite{Amendola:2010bk}, that unifies the first and second order formalisms for gravity theories. In C-theories, the connection is related to the conformally scaled
metric $\hat g_{\mu\nu}=C(\R)g_{\mu\nu}$ with a scaling that
depends on the scalar curvature $\R$. With nonlinear $f(\R)$,
C-theories interpolate and extrapolate the Einstein and Palatini
cases, and can avoid some of their conceptual and observational
problems. It is important to stress that a  bi-metric structure naturally comes out in this context by solving the Palatini equation for the connection  \cite{Olmoetal, allemandi} and the problem of the {\it dark side} of the Universe could be re-conducted to the confrontation of a metric in the Einstein frame to  another one in the Jordan frame \cite{dark}.

In an earlier work \cite{Flanagan:2003iw}, the known
equivalence between higher order gravity theories and scalar
tensor theories was generalized to a new class of theories. More
specifically, in the context of the Palatini formalism, where the
metric and connection are treated as independent variables (see
\cite{Olmo:2011uz} for a recent review), the Lagrangian density
was generalized to a function of the Ricci scalar, computed from
the metric, and a second Ricci scalar, computed from the
connection. These theories can be written as tensor-multi-scalar
theories, with two or more scalar fields. It is important to stress that such an approach select the whole budget of degrees of freedom due to the gravitational field in generalized theories of gravity. Essentially, dynamics of any scalar field is described by an effective Klein-Gordon  equation which comes out beside the standard Einstein field equations. In particular.  a fourth-order metric theory of gravity can be described as a second order Einstein theory plus a second order Klein-Gordon theory; a sixth-order theory is equivalent to Einstein plus two Klein-Gordon equations and so on. The results is that effective scalar field description can put in evidence all the gravitational degrees of freedom of a given theory\footnote{Recent progress shows that a nonperturbative approach, where one takes into account an infinite number of higher derivatives, results in qualitatively different theory that may even exhibit asymptotic freedom for gravity \cite{Biswas:2011ar}.} \cite{revnoi}.

More radically, one may modify the response of matter to gravity
by defining an action which depends nonlinearly upon the matter
Lagrangian \cite{Harko:2010mv}, or even its trace
\cite{Poplawski:2006ey}. Generally, the motion is non-geodesic,
and in fact, in these cases, the motion of matter is typically
altered already in flat Minkowski space, and one may expect
instabilities due to new nonlinear interactions within the matter
sector. A natural way to obtain solely gravitational modifications
of the behavior of matter emerges in the Palatini formulation of
extended gravity actions. In this approach, the relation between
the independent connection and the metric turns out to depend upon
the trace of the matter stress energy tensor in such a way that
the field equations effectively feature extra terms, given by the
matter content. However, since the extra terms contain fourth
order derivatives, the theory is problematical both at the
theoretical and phenomenological levels \cite{Olmo:2006zu,Koivisto:2005yc}.

In this paper, we will consider cosmological aspects of the hybrid 
metric-Palatini gravity considered in Ref. \cite{Harko:2011nh}.
In particular, starting from the field equations derived assuming as variable $X=\kappa^2T+R$, comprehensive of matter stress-energy tensor trace and the Ricci scalar, we obtain cosmological solutions, where matter and curvature can be discussed under the same footing. 

The present paper is organized as follows. In Sec. \ref{sect2}, we derive the field equations for $f(X)$ gravity in $D$-dimensions showing the general aspects by which metric and Palatini features mix by using $X$ as leading variable. Then we specialize to $D=4$ and derive the cosmological equations. Standard matter and curvature scalar can be used to derive effective pressure and matter-energy density.  The scalar-tensor representation of the theory is considered in Sec. \ref{sect3}. Here we show that $f(X)$-gravity can be recast as a scalar-tensor theory of gravity but the fact that we are using a hybrid metric-Palatini theory has important physical consequences that allow to avoid shortcomings that emerge if one takes into account pure metric or Palatini approaches. 
In Sec. \ref{sec:cosm} we derive the cosmological background equations, solve them in an example model and formulate the model as a dynamical system.
Section \ref{sect4} is devoted to the analysis of cosmological models emerging from $f(X)$-gravity in $4D$. We also go further to consider fluctuations around the background solution in Sec. \ref{sec:pert}, where we derive the cosmological perturbation equations and use them to show that the structure formation is free of instabilities, with however some new signatures that can be used to probe and test the models emerging from the new class of $f(X)$ gravity theories. Conclusions are drawn in Sec. \ref{sect5}.

\section{ $f(X)$-gravity}\label{sect2}

\subsection{The field equations in $D$-dimensions}

Let us start by considering  the action for hybrid metric-Palatini gravity in $D$-dimensions given by
 \be \label{action} S= \int d^D x \sqrt{-g} \lb
R + f(\R) + 2\ka^2 \mathcal{L}_m \rb\,, \ee
where, in addition to the
Einstein-Hilbert term and the matter Lagrangian which we assume to
have the standard form, there is an extra term depending on both the
metric and an independent connection $\hat{\Ga}$ through \be \R
\equiv  g^{\mu\nu}\R_{\mu\nu} \equiv g^{\mu\nu}\lp
\hat{\Gamma}^\alpha_{\mu\nu , \alpha}
       - \hat{\Gamma}^\alpha_{\mu\alpha , \nu} +
\hat{\Gamma}^\alpha_{\alpha\lambda}\hat{\Gamma}^\lambda_{\mu\nu} -
\hat{\Gamma}^\alpha_{\mu\lambda}\hat{\Gamma}^\lambda_{\alpha\nu}\rp\,.\label{r_def}
\ee The definition of the metric Ricci scalar $R$ is, as usual,
the above formula unhatted. By solving the equation of motion for
the connection, one finds that it is compatible with the metric
$F(\R)^{\frac{2}{D-2}}g_{\mu\nu}$ that is conformally related  to
$g_{\mu\nu}$, when the conformal factor is given by
\be F(\R)
\equiv \frac{df(\R)}{d\R}\,. \ee
This implies that the Palatini Ricci tensor is \ba \label{ricci1} \R_{\mu\nu}  =  R_{\mu\nu} +
\frac{D-1}{D-2}\frac{1}{F^2(\R)}F(\R)_{,\mu}F(\R)_{,\nu}
  - \frac{1}{F(\R)}\nabla_\mu F(\R)_{,\nu} -
\frac{1}{(D-2)}\frac{1}{F(\R)}g_{\mu\nu}\Box F(\R)\,. \ea 

Varying the action (\ref{action}) with respect to the metric, we obtain
\be \label{efe1} G_{\mu\nu} +
F(\R)\R_{\mu\nu}-\frac{1}{2}f(\R)g_{\mu\nu} = \ka^2 T_{\mu\nu}\,,
\ee
where the matter stress energy tensor is defined as usual,
 \be \label{memt}
 T_{\mu\nu} \equiv -\frac{2}{\sqrt{-g}} \frac{\delta
 (\sqrt{-g}\mathcal{L}_m)}{\delta(g^{\mu\nu})}.
 \ee
We can solve $\R$ from the trace of this field equation which
yields \be \label{trace1} \frac{D}{2}f(\R)-F(\R)\R =- \ka^2 T + \lp
\frac{D}{2}-1\rp R \equiv X\,. \ee Assuming that the form of the
function $f(\R)$ allows solution for the above equation, we can
express $\R$ algebraically in terms of $X$. The variable $X$ thus
measures the failure of the theory to satisfy the general
relativistic trace equation which, for $D=4$, gives $R=-\ka^2
T$. We can then express the field equation (\ref{efe1}) in terms of
the metric and $X$ as \ba \label{efex} G_{\mu\nu} & = &
\frac{1}{2}f(X)g_{\mu\nu}- F(X)R_{\mu\nu}  +  F'(X)
\nabla_{\mu}X_{,\nu}
     + \frac{1}{D-2}\lb F'(X)\Box X + F''(X)\lp \partial X\rp^2 \rb
g_{\mu\nu}
    \nonumber  \\
&&+ \lb F''(X)-\frac{D-1}{D-2}\frac{\lp
F'(X)\rp^2}{F(X)}\rb X_{,\mu}X_{,\nu} + \ka^2 T_{\mu\nu}\,. \ea

Note that $(\partial X)^2=X_{,\mu}X^{,\mu}$. Due to the second
order derivatives acting upon $X$, the theory is fourth order in
derivatives of the metric. Note that here $F(X)$ is the short-hand
for $F(\R(X))$, and $F'(X)$ denotes $F'(X)=\partial
F(\R(X))/\partial X$, and so on. The trace of the field equations
yields
\ba \label{trace21} F'(X)\Box X + \lb
F''(X)-\frac{1}{2}\frac{\lp F'(X)\rp^2}{F(X)}\rb \lp \partial
X\rp^2
+ \frac{D-2}{2(D-1)}\left [ X + \frac{D}{2}f(X)-F(X)R\right]= 0 \,.
\ea
Note that the Ricci scalar $R$, computed from the metric, is related
to the Palatini $\R$ via 
\be \label{ricciscalar1} \R(X) =
R+\frac{D-1}{D-2}\lb \lp\frac{F'(X)}{F(X)}\rp^2-2\frac{\Box
F(X)}{F(X)}\rb\,, 
\ee 
which is seen by contracting the gravitational field equation, given by Eq. (\ref{ricci1}).
Thus Eq.~(\ref{trace21}) and Eq.~(\ref{ricciscalar1}) are redundant with
Eq.~(\ref{trace1}). Let us now specify the theory in 4-dimensions, which will be explored throughout this work.

\subsection{The gravitational field equations in 4-dimensions}

The above $D$-dimensional action can be specified as
\begin{equation} \label{eq:S_hybrid}
S= \frac{1}{2\kappa^2}\int d^4 x \sqrt{-g} \left[ R + f(\R)\right] +S_m \ ,
\end{equation}
where $S_m$ is the matter action, $\kappa^2\equiv 8\pi G$, $R$ is
the Einstein-Hilbert term, $\R  \equiv g^{\mu\nu}\R_{\mu\nu} $ is
the Palatini curvature, and $\R_{\mu\nu}$ is defined in terms of
an independent connection $\hat{\Gamma}^\alpha_{\mu\nu}$  as
\begin{equation}
\R_{\mu\nu} \equiv \hat{\Gamma}^\alpha_{\mu\nu ,\alpha} -
\hat{\Gamma}^\alpha_{\mu\alpha , \nu} +
\hat{\Gamma}^\alpha_{\alpha\lambda}\hat{\Gamma}^\lambda_{\mu\nu}
-\hat{\Gamma}^\alpha_{\mu\lambda}\hat{\Gamma}^\lambda_{\alpha\nu}\,.
\end{equation}
Varying the action given by Eq. (\ref{eq:S_hybrid}) with respect to the metric, one obtains the following gravitational field equation  
\be
\label{efe} G_{\mu\nu} +
F(\R)\R_{\mu\nu}-\frac{1}{2}f(\R)g_{\mu\nu} = \ka^2 T_{\mu\nu}\,.
\ee
The matter stress-energy tensor, as before, is defined by Eq. (\ref{memt}). Note that the
independent connection is compatible with the metric
$F(\R)g_{\mu\nu}$, conformal to $g_{\mu\nu}$, with the conformal\footnote{If the gravitational lagrangian depended upon other curvature invariants
besides $\R$, one would obtain a disformal relation between the two metrics \cite{Olmoetal}.}
factor  given by $F(\R) \equiv df(\R)/d\R$. This implies that
\ba
\label{ricci} \R_{\mu\nu} & = & R_{\mu\nu} +
\frac{3}{2}\frac{1}{F^2(\R)}F(\R)_{,\mu}F(\R)_{,\nu}
  - \frac{1}{F(\R)}\nabla_\mu F(\R)_{,\nu} -
\frac{1}{2}\frac{1}{F(\R)}g_{\mu\nu}\Box F(\R)\,. \ea 
The Palatini curvature, $\R$, can be
obtained from the trace of the field equation (\ref{efe}), which
yields \be \label{trace} F(\R)\R -2f(\R)= \ka^2 T +  R \equiv X\,.
\ee
 As above, we can express $\R$ algebraically in
terms of $X$ if the form of $f(\R)$ allows analytic solutions. Again, the variable $X$ measures how much  the theory deviates from the
general relativity trace equation $R=-\ka^2 T$. The field equation
(\ref{efe}) can be recast as $G_{\mu\nu}=\ka^2
T^{\rm eff}_{\mu\nu}$, where $ T^{\rm eff}_{\mu\nu}= T^{\rm
X}_{\mu\nu}+T_{\mu\nu}$, with $T^{\rm X}_{\mu\nu}$ defined as
\ba
\label{Tefex} T^{\rm X}_{\mu\nu}& = &\frac{1}{\ka^2} \Bigg\{
\frac{1}{2}f(X)g_{\mu\nu}- F(X)R_{\mu\nu}  +  F'(X)
\nabla_{\mu}X_{,\nu}
     + \frac{1}{2}\lb F'(X)\Box X + F''(X)\lp \partial X\rp^2 \rb
g_{\mu\nu}
    \nonumber  \\
&&+ \lb F''(X)-\frac{3}{2}\frac{\lp F'(X)\rp^2}{F(X)}\rb
X_{,\mu}X_{,\nu} \Bigg\}\,. \ea
 The trace of the
field equations is now \ba \label{trace2} F'(X)\Box X + \lb
F''(X)-\frac{1}{2}\frac{\lp F'(X)\rp^2}{F(X)}\rb \lp \partial
X\rp^2+
 \frac{1}{3}\left[ X + 2f(X)-F(X)R\right]= 0 \,,
\ea  while the relation between the metric  scalar curvature  $R$ and
the Palatini scalar curvature  $\R$ is   
\be \label{ricciscalar} \R(X) =
R+\frac{3}{2}\lb \lp\frac{F'(X)}{F(X)}\rp^2-2\frac{\Box
F(X)}{F(X)}\rb\,, \ee 
which can be obtained by contracting
Eq.~(\ref{ricci}).  Now we have all the ingredients to derive cosmological equations from this theory.

\subsection{Scalar-tensor representation of $f(X)$-gravity}\label{sect3}

Like in the pure metric and Palatini cases \cite{Olmo:2005jd}, the action (\ref{eq:S_hybrid}) for $f(X)$ theories can be turned into that of a
scalar-tensor theory by introducing an auxiliary field $A$ such
that
\begin{equation} \label{eq:S_scalar0}
S= \frac{1}{2\kappa^2}\int d^4 x \sqrt{-g} \left[ R + f(A)+f_A(\R-A)\right] +S_m \ ,
\end{equation}
where $f_A\equiv df/dA$. Rearranging the terms and defining $\phi\equiv f_A$, $V(\phi)=A f_A-f(A)$,
Eq. (\ref{eq:S_scalar0}) becomes
\begin{equation} \label{eq:S_scalar1}
S= \frac{1}{2\kappa^2}\int d^4 x \sqrt{-g} \left[ R + \phi\R-V(\phi)\right] +S_m \ .
\end{equation}
It is important to stress that in this case we are also considering a hybrid metric-Palatini theory.
Variation of this action with respect to the metric, the scalar $\phi$ and the connection leads to the field equations
\begin{eqnarray}
R_{\mu\nu}+\phi \R_{\mu\nu}-\frac{1}{2}\left(R+\phi\R-V\right)g_{\mu\nu}&=&\kappa^2 T_{\mu\nu}\,,
\label{eq:var-gab}\\
\R-V_\phi&=&0 \label{eq:var-phi} \,, \\
\hat{\nabla}_\alpha\left(\sqrt{-g}\phi g^{\mu\nu}\right)&=&0 \,, \label{eq:connection}\
\end{eqnarray}
respectively. 

The solution of Eq.~(\ref{eq:connection}) implies that the independent connection is the Levi-Civita connection of
a metric $h_{\mu\nu}=\phi g_{\mu\nu}$. This means that we are dealing with a bi-metric theory and $\R_{\mu\nu}$ and $R_{\mu\nu}$ are related by
\begin{equation} \label{eq:conformal_Rmn}
\R_{\mu\nu}=R_{\mu\nu}+\frac{3}{2\phi^2}\partial_\mu \phi \partial_\nu \phi-\frac{1}{\phi}\left(\nabla_\mu
\nabla_\nu \phi+\frac{1}{2}g_{\mu\nu}\Box\phi\right) \ ,
\end{equation}
which can be used in the action (\ref{eq:S_scalar1}) to get rid of the independent connection and obtain the following
scalar-tensor representation that belongs to the ``family of scalar-tensor theories'' \cite{Koivisto:2009jn}, so that we finally arrive at the following action 
\begin{equation} \label{eq:S_scalar2}
S= \frac{1}{2\kappa^2}\int d^4 x \sqrt{-g} \left[ (1+\phi)R +\frac{3}{2\phi}\partial_\mu \phi \partial^\mu \phi
-V(\phi)\right] +S_m \ .
\end{equation}
It is important to point out that, by the substitution $\phi \rightarrow -(\kappa\phi)^2/6$, the action (\ref{eq:S_scalar2}) reduces to the well-known case of a conformally coupled scalar field with a self-interaction potential. Precisely, this redefinition makes the kinetic term in the action (\ref{eq:S_scalar2}) the standard one, and the action itself becomes that of a massive scalar-field conformally coupled to the Einstein gravity.  Of course, it is not the Brans-Dicke gravity where the scalar field is massless.

As we will see, this simple modification will have important physical
consequences. Thus, we have made contact with the general class of scalar-tensor  theories including  nonlinear couplings and  potentials. Such theories have been previously considered in cosmology, in particular a reconstruction method to deduce the coupling and the potential form observations of large scale structure and expansion history was outlined already in \cite{Boisseau:2000pr}. More recent studies have delved into such issues as screening phenomena and nonlinear structure formation, see e.g. \cite{stt}. Here we are motivated to
study the particular class of such actions: remarkably, we have arrived at this class of theories, that  was already single out by the consideration of an algebraic property \cite{Koivisto:2009jn}, here from a quite different physical starting point writing down the maybe simplest action incorporating both the metric and an independent connection.

Using Eq.~(\ref{eq:conformal_Rmn}) and Eq.~(\ref{eq:var-phi}) in Eq.~(\ref{eq:var-gab}), the metric field equation can be
written as
\begin{eqnarray}
(1+\phi) R_{\mu\nu}&=&\kappa^2\left(T_{\mu\nu}-\frac{1}{2}g_{\mu\nu} T\right)+\frac{1}
{2}g_{\mu\nu}\left(V+\Box\phi\right)+\nabla_\mu\nabla_\nu\phi-\frac{3}{2\phi}\partial_\mu \phi
\partial_\nu \phi \ \label{eq:evol-gab} ,
\end{eqnarray}
from which it follows that the spacetime curvature is generated by both the matter and the scalar field.
The scalar field equation can be manipulated in two different ways that illustrate how this theory is related
with the $w=0$ and $w=-3/2$ cases, which corresponds to the metric and Palatini scalar-tensor representations of $f(R)$-gravity \cite{revnoi}, respectively. Tracing Eq.~(\ref{eq:var-gab}) with $g^{\mu\nu}$,  we find $-
R-\phi\R+2V=\kappa^2T$, and using Eq.~(\ref{eq:var-phi}), takes the following form
\begin{equation}\label{eq:phi(X)}
2V-\phi V_\phi=\kappa^2T+R \ .
\end{equation}

Similarly as in the Palatini ($w=-3/2$) case, this equation tells
us that the field $\phi$ can be expressed as an algebraic function
of the scalar $X\equiv \kappa^2T+R$, i.e., $\phi=\phi(X)$. In the
pure Palatini case, however, $\phi$ is just a function of $T$. The
right-hand side of Eq.~(\ref{eq:evol-gab}), therefore, besides
containing new matter terms associated with the trace $T$ and its
derivatives, also contains the curvature $R$ and its derivatives.
Thus, this theory can be seen as a higher-derivative theory in
both  matter and  metric fields. However, such an
interpretation can be avoided if $R$ is replaced in
Eq.(\ref{eq:phi(X)}) with the relation $R=\R+\frac{3}{\phi}\Box
\phi-\frac{3}{2\phi^2}\partial_\mu \phi \partial^\mu \phi$
together with $\R=V_\phi$. One then finds that the scalar field is
governed by the second-order evolution equation
\begin{equation}\label{eq:evol-phi}
-\Box\phi+\frac{1}{2\phi}\partial_\mu \phi \partial^\mu
\phi+\frac{\phi[2V-(1+\phi)V_\phi]} {3}=\frac{\phi\kappa^2}{3}T\,,
\end{equation}
which is an effective Klein-Gordon equation.
This last expression shows that, unlike in the Palatini ($w=-3/2$)
case, the scalar field is dynamical. The theory is therefore not
affected by the microscopic instabilities that arise in Palatini
models with infrared corrections \cite{Olmo:2006zu,Olmo:2011uz}.

With these considerations in mind, we shall consider $f(X)$-cosmology in the scalar-tensor representation below.

\section{$f(X)$-cosmological field equations}
\label{sec:cosm}

\subsection{$f(X)$ gravity and the late-time cosmic acceleration}

As a specific example of $f(X)$
cosmological dynamics, let us  consider the  spatially flat
Friedman-Robertson- Walker (FRW) metric given by the metric element
\begin{equation}
ds^2=-dt^2+a^2(t) d{\bf x}^2 \,,
\end{equation}
where $a(t)$ is the scale factor. The Ricci scalar is given by
$R=6(2H^2+\dot{H})$, where $H=\dot{a}(t)/a(t)$ is the Hubble
parameter, and the overdot denotes a derivative with respect to
the cosmic time. As usual, the standard matter conservation
law for the energy density $\rho_{m}$ and pressure $p_{m}$ is given by
$\dot{\rho}_{m}+3H(\rho_{m}+p_{m})=0$.

Thus, the modified Friedmann equations take the following form:
\ba
\kappa^2\rho_{\rm eff}&=&3H^2 \\
\kappa^2 p_{\rm eff}&=&-(2\dot{H}+3H^2)\,, 
\ea
where $\rho_{\rm eff}$ and $p_{\rm eff}$
are the total effective energy density and pressure, respectively,
and are given by
\begin{eqnarray}
\rho_{\rm
eff}&=&\rho-\frac{1}{\kappa^2}\Bigg\{\frac{1}{2}f(X)-\frac{3}{2}
\left[F''(X)-\frac{(F'(X))^2}{F(X)}\right]\dot{X}^2
     -\frac{3}{2}F'(X)\left(\ddot{X}+\dot{X}H\right)
 -3F(X)\left(\dot{H}+H^2 \right)\Big\}
\,,   \label{rhoeff} \\
p_{\rm eff}&=&p+\frac{1}{\kappa^2}\Bigg\{\frac{1}{2}f(X)-\frac{1}{2}F''(X)\dot{X}^2
     -\frac{1}{2}F'(X)\left(\ddot{X}+5\dot{X}H\right)
 -F(X)\left(\dot{H}+3H^2 \right)\Big\}\,,
 \label{peff}
\end{eqnarray}
respectively.
For an effective equation of state defined by the parameter
$w_{\rm eff}=p_{\rm eff}/\rho_{\rm eff}$, we obtain $w_{\rm
eff}=-1-2\dot{H}/3H^2$.
The late-time cosmic acceleration occurs when the strong energy
condition (SEC) is violated, i.e., $\rho_{\rm eff}+3p_{\rm eff}<0$
or $w_{\rm eff}<-1/3$. For simplicity, considering the specific
case of vacuum $\rho_{m}=p_{m}=0$, the SEC is given by
\begin{eqnarray}
\rho_{\rm eff}+3p_{\rm eff}&=&\frac{1}{\kappa^2}
\left\{f(X)-6H\left[F'(X)\dot{X}+FH \right]
   -\frac{3}{2}\dot{X}\frac{(F'(X))^2}{F(X)}\right\}  \,.
\end{eqnarray}

As a toy model, let us consider the power law models
$f(\R)=\al \R^n$, where $\alpha $ and $n$ constant parameters. Using
Eq.~(\ref{trace}), we obtain \ba
f(X) & = & \frac{X}{n-2}  \nonumber \\
F(X) &=&
n\al\lb\frac{X}{\al (n-2)}\rb^{1-\frac{1}{n}}\,, \nonumber \\
F'(X) & = & \frac{n-1}{n-2}\lb\frac{X}{\al (n-2)}\rb^{-\frac{1}{n}}\,,
\nonumber  \\
F''(X)&=&\frac{-(n-1)}{\al n(n-2)^2}\lb\frac{X}{\al (n-2)}
\rb^{-1-\frac{1}{n}}\,. \nonumber \ea 
As in the pure $f(\R)$ or
``Palatini'' gravity, the case $n=2$ is degenerate, and in the
following we assume $n \neq 2$.
The general condition for the acceleration of the Universe can be
formulated as
\begin{eqnarray}
36H^2n^2X^2 + 9\left(n -1  \right) \left( n-1  + 4HnX \right)\dot{X}>
\frac{1}{3}nX\left(6X+2n-4\right){\left[ \frac{X}{\left( n-2
\right) \alpha} \right] }^{1/n}.
\end{eqnarray}
To look for power-law solutions for the scale factor, $a \sim t^\beta$, we 
plug such an ansatz into the Eqs.~(\ref{rhoeff})-(\ref{peff}). 

For generic $n$, we find that the consistent solutions
are described by the effective equation of state parameter, $w_{\rm eff}=p_{\rm eff}/\rho_{\rm eff}$, given by
\ba \label{weff}
w_{\rm eff}& =&-1+ \frac{4\beta \left[ 1+4\beta(1-\beta)-n\beta(4\beta^2-8\beta+5)   \right]}{f(n,\alpha,\beta,t)+6n\beta \left[ n\beta (4\beta^3-12\beta^2+13\beta-6) +\beta (4\beta^2-8\beta+5) +n-1 \right]} = -1+\frac{2}{3}\beta\,,
\ea
where, for notational simplicity, $f(n,\alpha,\beta,t)$ is defined as
\be
f(n,\alpha,\beta,t)= n\alpha (n-2)(2\beta -1)t^46^{1/n}\left[\frac{\beta (2\beta -1)}{t^2 \alpha (n-2)} \right]^{\frac{n+1}{n}} \,.
\ee
The first equality in Eq. (\ref{weff}) follows from inserting the ansatz in the field equations, the second one directly from the ansatz.
One may now take specific values for the parameters $\alpha$ and $n$ to solve away $\beta$ and obtain $w_{\rm eff}<-1/3$ to predict an accelerated behavior of the Hubble fluid.

As a specific example, for the case of $n=-1$, so that the effective equation of state parameter, $w_{\rm eff}$ simplifies to
\ba
w_{\rm eff}=-\frac{6\beta^3-25\beta^2+31\beta -10 -\frac{\alpha}{\beta}t^4}{6\beta^3-21\beta^2+21\beta -6 -\frac{\alpha}{\beta}t^4} = -1+\frac{2}{3}\beta  \,.
\ea
One of the solutions is the de Sitter expansion,
\be
w_{\rm eff} = -1\,,
\ee
thus we expect that in the $f \sim 1/\R$ model the universe ends up in an accelerating phase.

\subsection{General dynamical system analysis}

To explore the general dynamical analysis, consider the $f(X)$-gravity action given by
\be \label{s-t}
S=\frac{1}{2\kappa^2}\int d^4 x \sqrt{-g}\lb \OA R + f(\R)\rb + S_m
\ee
which is equivalent to the scalar tensor theory of the so called algebraic class\footnote{By algebraic we mean that there always exists a functional relation $\phi=\phi(\OA R + \kappa^2T)$, but we stress that the scalar field is nevertheless dynamical except in the special case $\OA=0$ where one recovers the Palatini-$f(\R)$ theory \cite{Koivisto:2009jn}.}. Note that we include the parameter $\OA$ here for generality, and then the result in the previous section becomes 
\be
S=\frac{1}{2\kappa^2}\int d^4 x \sqrt{-g}\lb (\OA+\phi) R + \frac{3}{2\phi}\lp \partial\phi\rp^2 - 2\kappa^2V(\phi)\rb + S_m\,,
\ee
where
\be
\kappa^2V(\phi)=\frac{1}{2}\lb r(\phi)\phi-f(r(\phi))\rb, \quad r(\phi) \equiv {f'}^{-1}(\phi) \,.
\ee

The Friedmann equations can always be written in terms of the effective energy density and pressure, respectively defined as
\ba
3H^2 & = & \kappa^2\rho_{\rm eff} - \frac{K}{a^2}\,, \label{fr1} \\
\dot{H} & = & -\frac{\kappa^2}{2}\lp\rho_{\rm eff}+p_{\rm eff}\rp + \frac{K}{a^2} \label{fr2} \,.
\ea
For simplicity, we consider the flat universe $K=0$ in the following. For the theory provided by the action (\ref{s-t}) we obtain the following modified Friedmann equations
\ba
\lp \OA+\phi \rp \kappa^2\rho_{\rm eff} & = & -\frac{3}{4\phi}\dot{\phi}^2 + \kappa^2V(\phi) - 3H\dot{\phi} + \kappa^2\rho_m\,, \\
\lp \OA+\phi \rp \kappa^2 p_{\rm eff} & = & -\frac{3}{4\phi}\dot{\phi}^2 - \kappa^2V(\phi) + \ddot{\phi} + 2H\dot{\phi} + \kappa^2 p_m\,,
\ea
respectively.

The conservation equations for the matter component and the scalar field are
\ba
\dot{\rho}_m + 3H(\rho_m + p_m) & = & 0\,, \\
\ddot{\phi} + 3H\dot{\phi} - \frac{\dot{\phi}^2}{2\phi} + \frac{1}{3}\phi R - \frac{2}{3}\kappa^2 \phi V'(\phi) & = & 0 \label{kg}\,.
\ea
Recalling that $R=6(2H^2+\dot{H})$ and using Eqs. (\ref{fr1}) and (\ref{fr2}), we can rewrite the Klein-Gordon equation as
\be
\ddot{\phi}+3H\dot{\phi} - \frac{\dot{\phi}^2}{2\phi} + U'(\phi) + \frac{\kappa^2 \phi}{3\OA}\lp \rho_m-3p_m\rp =0\,, \label{kg2}
\ee
where for notational simplicity, $U'(\phi)$ is defined by
\be
U'(\phi) \equiv \frac{2\kappa^2\phi}{3\OA}\lb 2V(\phi) - \lp \OA + \phi \rp V'(\phi)\rb\,. \label{veff}
\ee
As a consistency check one can verify that the Klein-Gordon equation together with the matter conservation allows to derive (\ref{fr2}) from (\ref{fr1}).
By combining Eqs. (\ref{kg}) and (\ref{kg2}), we find that
\be
2V(\phi)-V'(\phi)\phi =\frac{1}{2}\lp \OA R + \kappa^2 T_m \rp \equiv \frac{1}{2}X\,. \label{alg}
\ee
The solution for $\phi=\phi(X=0)$ gives us the natural initial condition for the field in the early universe. The asymptotic value of the field in the far future may then be deduced by studying the minima of the function $U(\phi)$ defined by Eq. (\ref{veff}).

In order to study the dynamical system, we introduce the dimensionless variables
\be
\Omega_m  \equiv  \frac{\kappa^2\rho_m}{3H^2}\,, \quad
x  \equiv  \phi\,, \quad y = x_{,N}\,, \quad z = \frac{\kappa^2V}{3H^2}\,,
\ee
where $N=\log{a}$ is the e-folding time. The Friedmann equation (\ref{fr1}) can then be rewritten as
\be
\OA + x + y -z +\frac{y^2}{4x} = \Omega_m\,.
\ee
Due to this constraint, the number of independent degrees of freedom is three instead of four. We choose to span our phase space by the triplet $\{x,y,z\}$. The autonomous system of equations for them reads as
\ba
x_{,N} & = & y\,, \\
y_{,N} & = & \frac{2x+y}{8\OA x}\Big\{ \lp 3w_m-1 \rp y^2 + 4x\lb \lp 3w_m-1 \rp y-3\lp 1+w_m\rp z\rb \nonumber \\
&& - 4x^2\lp 1-3w_m-2u(x)z\rp +4 \OA\lb 3x\lp w_m-1\rp y+y^2-x^2\lp 2-6w_m-4u(x)z\rp\rb\Big\}\,, \\
z_{,N} & = & \frac{z}{4\OA x}\Big\{\lp 3w_m-1\rp y^2  +  4x\lb\lp 3w_m-1\rp y - 3\lp 1+w_m\rp z\rb
\nonumber \\ 
&& + 4\OA x \lp 3+3w_m+u(x)y\rp + 4x^2\lp 3w_m-1+2u(x)z\rp\Big\}\,,
\ea
respectively.
We have defined $u(x) \equiv V'(\phi)/V(\phi)$. The relevant fixed points appear in this system. In particular, we have the matter dominated fixed point where $x=y=z=0$ and $w_{\rm eff}=w_m$, and the de Sitter fixed point\footnote{In addition, there exists the fixed point $x=-\OA$ corresponding to some kind of singular evolution.} that is described by $w_{\rm eff}=-1$ and
\be
x_*=(2-\OA u_*)/u*\,, \quad y_*=0\,, \quad z_* = 2/u_*\,. \label{dS}
\ee
We denote the asymptotic values corresponding to this fixed point by a subscript star. In particular, the asymptotic value of the field $x_*$ is solved from the first equation in (\ref{dS}) once the form of the potential is given. As expected, this value corresponds to minimum of the effective potential (\ref{veff}), $U'(x_*)=0$. To construct a viable model, the potential should be such that we meet the two requirements:
\begin{itemize}
\item The matter dominated fixed point should be a saddle point, the de Sitter fixed point an attractor. Then we naturally obtain a transition to acceleration following standard cosmological evolution.
\item At the present epoch the field value should be sufficiently close to zero. Then we avoid conflict with the Solar system tests of gravity.
\end{itemize}
Note that the simplest metric $f(R)$ theories that provide acceleration utterly fail in {\it both} predicting a viable structure formation era and the Solar system as we observe it. The Palatini-$f(\R)$ models on the other hand can be ruled out as dark energy alternative by considering their structure formation or implications to microphysics, if such a theory is regarded consistent in the first place. As shown here and explored further below, $f(X)$ gravity models exist that are free of these problems.
\newline
\newline
To summarize: the field goes from $\phi_i$ to $\phi_*$, where the former is given by $2V(\phi_i)=V'(\phi_i)\phi_i$ and the latter by $2V(\phi_*)=(\OA+\phi_*)V'(\phi_*)$. We just need a suitable function $V(\phi)$, i.e. $f(\R)$ in such a way that the slope will be downwards and $\phi_*$ near the origin.

\section{Cosmology of $f(X)$-gravity: analytical solutions}\label{sect4}

As we have seen, the above theory of $f(X)$-gravity can be recast in terms of a non-minimally coupled scalar-tensor  gravity. The scalar field represents the ``deviation'' of the theory from the standard general relativity where standard matter is also considered. In particular, the auxiliary variable $X$ states ``how much'' the trace equation of the theory deviates from the general relativistic one. An important point to stress is that such a scalar field has a purely geometric origin and describes  further degrees of freedom of the gravitational field coming from  extended theories of gravity.

It is straightforward to rewrite the cosmological equations in the absence of standard matter as
\begin{eqnarray}
H^{2}&=&\frac{\kappa ^{2}}{3}\rho _{\rm eff},\\
\dot{H}&=&-\frac{\kappa ^{2}}{2}\left(
\rho _{\rm eff}+p_{\rm eff}\right) ,
\end{eqnarray}
where
\begin{eqnarray}
\left( 1+\phi \right) \kappa ^{2}\rho _{\rm eff}&=&-\frac{3}{4\phi }\dot{\phi}%
^{2}+\kappa ^{2}V\left( \phi \right) -3H\dot{\phi},  \label{1a}
\\
\left( 1+\phi \right) \kappa ^{2}p_{\rm eff}&=&-\frac{3}{4\phi }\dot{\phi}%
^{2}-\kappa ^{2}V\left( \phi \right) +2H\dot{\phi}+\ddot{\phi}.
\label{2a}
\end{eqnarray}
It is important to stress that there is no standard matter in the definition of pressure and energy density, however, due to the definition of the scalar field $\phi$ and its dependence on $X$ curvature behaves as a perfect fluid.
The scalar field  satisfies the Klein-Gordon equation,
\begin{equation}
\ddot{\phi}+3H\dot{\phi}-\frac{1}{2\phi }\dot{\phi}^{2}+\frac{2\kappa ^{2}}{3%
}\phi \left[ 2V\left( \phi \right) -\left( 1+\phi \right)
V^{\prime }\left( \phi \right) \right] =0.  \label{3a}
\end{equation}
Note that as an indicator of the accelerated expansion one can consider the
behavior of the deceleration parameter, given by
\begin{equation}
q=\frac{d}{dt}\frac{1}{H}-1=-\frac{\dot{H}}{H^{2}}-1=\frac{3}{2}\frac{\rho
_{\rm eff}+p_{\rm eff}}{\rho _{\rm eff}}-1.  \label{4a}
\end{equation}
where accelerated expansion occurs when $q<0$. 

From Eqs.~(\ref{1a}) and (\ref{2a}), we obtain
\begin{equation}
\left( 1+\phi \right) \kappa ^{2}\left( \rho _{\rm eff}+p_{\rm eff}\right) =-\frac{3%
}{2\phi }\dot{\phi}^{2}-H\dot{\phi}+\ddot{\phi}.  \label{5a}
\end{equation}
By eliminating $\ddot{\phi}$ with the help of the Klein-Gordon Eq.~(\ref{3a}), we get
\begin{equation}
\left( 1+\phi \right) \kappa ^{2}\left( \rho _{\rm eff}+p_{\rm eff}\right) =-\frac{1}{\phi }\dot{\phi}^{2}-4H\dot{\phi}-\frac{2\kappa ^{2}}{3}\phi
\left[
2V\left( \phi \right) -\left( 1+\phi \right) V^{\prime }\left( \phi \right) %
\right] .  \label{6a}
\end{equation}
Therefore the cosmological equations become
\begin{eqnarray}
3H^{2}&=&\frac{1}{1+\phi }\left[ -\frac{3}{4\phi }\dot{\phi}^{2}-3H\dot{\phi}%
+\kappa ^{2}V\left( \phi \right) \right] ,  \label{6b}
 \\
2\dot{H}&=&\frac{1}{1+\phi }\left[ \frac{1}{\phi }\dot{\phi}^{2}+4H\dot{\phi}+%
\frac{2\kappa ^{2}}{3}\phi \left[ 2V\left( \phi \right) -\left(
1+\phi \right) V^{\prime }\left( \phi \right) \right] \right] .
\label{6c}
\end{eqnarray}
The explicit dependence on the scalar field for $q$ is
\begin{equation}
q=-\frac{3}{2}\left\{\frac{\dot{\phi}^{2}/\phi +4H\dot{\phi}+2\kappa
^{2}\phi \left[
2V\left( \phi \right) -\left( 1+\phi \right) V^{\prime }\left( \phi \right) %
\right] /3}{-3\dot{\phi}^{2}/4\phi +\kappa ^{2}V\left( \phi \right) -3H\dot{%
\phi}}\right\}-1\,.  \label{7a}
\end{equation}
Clearly, the dynamics of cosmological models can be classified according to Eq.~(\ref{7a}). Reversing the argument, conditions on Eq.~(\ref{7a}) assign, in principle, the functional form of $f(X)$.

\subsection{Marginally accelerating models}

A particular class of models, which may be called ``marginally
accelerating'', are those satisfying the condition $q=0$, which
gives the equation
\begin{equation}
\dot{\phi}^{2}/\phi +4H\dot{\phi}+2\kappa ^{2}\phi \left[ 2V\left(
\phi
\right) -\left( 1+\phi \right) V^{\prime }\left( \phi \right) \right] /3+%
\frac{2}{3}\left[ -3\dot{\phi}^{2}/4\phi +\kappa ^{2}V\left( \phi \right) -3H%
\dot{\phi}\right] =0,  \label{8}
\end{equation}
and then
\begin{equation}
\frac{\dot{\phi}^{2}}{2\phi }+2H\dot{\phi}+\frac{2}{3}\kappa
^{2}\phi \left[
3V\left( \phi \right) -\left( 1+\phi \right) V^{\prime }\left( \phi \right) %
\right] =0.
\end{equation}

The simplest marginally accelerating model can be obtained by
assuming that the potential satisfies the condition
\begin{equation}
3V\left( \phi \right) -\left( 1+\phi \right) V^{\prime }\left(
\phi \right) =0,
\end{equation}
which yields the following solution 
\begin{equation}
V(\phi )=V_{0}\left( 1+\phi \right) ^{3},
\end{equation}%
with $V_{0}$ an arbitrary constant of integration. We also obtain
immediately
\begin{equation}
\phi =\frac{\phi _{0}}{a^{4}},\qquad H=-\frac{\dot{\phi}}{4\phi }.
\end{equation}

The  Friedmann equation (\ref{6a}) becomes
\begin{equation}
\frac{3}{16}\frac{\dot{\phi}^{2}}{\phi ^{2}}=\kappa
^{2}V_{0}\left( 1+\phi \right) ^{2},
\end{equation}
from which one obtains the solutions
\begin{equation}
\phi \left( t\right) =\frac{\exp \left[ -4\kappa \sqrt{\frac{V_{0}}{3}}%
\left( t-t_{0}\right) \right] }{1-\exp \left[ -4\kappa \sqrt{\frac{V_{0}}{3}}%
\left( t-t_{0}\right) \right] },
\end{equation}
and
\begin{equation}
a(t)=a_{0}\left\{ \exp \left[ 4\kappa \sqrt{\frac{V_{0}}{3}}\left(
t-t_{0}\right) \right] -1\right\} ^{1/4}.  \label{9a}
\end{equation}

The importance of the marginally accelerating models is due to the
fact that since the deceleration parameter is a monotonically
decreasing function of time, models that reached the value $q=0$
starting from decelerating states with $q>0$ will end in an
accelerating state with $q<0$, as in fact can be  seen from
Eq.~(\ref{9a}). In this way we have established the existence of at
least one accelerating solution in the $f(X)$-gravity model. Furthermore, these types of solutions acquire a physical meaning being the possible junction between any dust-dominated era $(q>0)$ and recent accelerated expansion $(q<0)$. In some sense, they are the turning point between an epoch of structure formation  and dark energy.

\subsection{Accelerating models}

\subsubsection{General considerations}

To tackle, in general, the problem of accelerating models, let us
 simplify the mathematical formalism by introducing the
new variable $x=\ln a$, $dx/dt=\dot{a}/a=H$. Then we have
\begin{equation}
\frac{d}{dt}=\frac{d}{dx}\frac{dx}{dt}=\frac{d}{dx}\frac{\dot{a}}{a}=H\frac{d%
}{dx},
\end{equation}
and
\begin{equation}
\frac{d^{2}}{dt^{2}}=\dot{H}\frac{d}{dx}+H\frac{d^{2}}{dx^{2}}\frac{dx}{dt}=%
\dot{H}\frac{d}{dx}+H^{2}\frac{d^{2}}{dx^{2}}.
\end{equation}

In these variables, the first modified Friedmann equation (\ref{6b}) becomes
\begin{equation}
H^{2}=\frac{\kappa ^{2}}{3}\left[\frac{V\left( \phi \right) }{1+\phi
+\phi _{,x}+\phi _{,x}^{2}/4\phi }\right].\label{pippo}
\end{equation}
On the other hand, the second Friedmann equation (\ref{6c}) takes the form
\begin{equation}
\frac{\dot{H}}{H^{2}}=\frac{1}{2(1+\phi) }\left\{ \frac{1}{\phi }%
\phi _{,x}^{2}+4\phi _{,x}+\frac{2\kappa ^{2}}{3}\frac{\phi \left[
2V\left(
\phi \right) -\left( 1+\phi \right) V^{\prime }\left( \phi \right) \right] }{%
H^{2}}\right\} ,
\end{equation}
and the deceleration parameter is provided by
\begin{equation}
q=-\frac{1}{2(1+\phi) }\left\{ \frac{1}{\phi }\phi
_{,x}^{2}+4\phi _{,x}+2\frac{\phi \left[ 2V\left( \phi \right)
-\left( 1+\phi \right) V^{\prime }\left( \phi \right) \right]
\left( 1+\phi +\phi _{,x}+\phi _{,x}^{2}/4\phi \right) }{V(\phi
)}\right\} -1.
\end{equation}%
This result allows to formulate  the condition for accelerating expansion  as
\begin{equation}
\frac{V_{,x}}{V}<\frac{\left[ 2\left( 1+\phi \right) +\phi
_{,x}^{2}/\phi +4\phi _{,x}\right] \phi _{,x}}{2\phi \left( 1+\phi
\right) \left( 1+\phi +\phi _{,x}+\phi _{,x}^{2}/4\phi \right)
}+\frac{2\phi _{,x}}{1+\phi },
\end{equation}%
where we have used $dV/d\phi =\left( dV/dx\right) \left( dx/d\phi
\right) =\left( dV/dx\right) \left( 1/\phi _{,x}\right) $. The
Klein-Gordon equation (\ref{3a}) can be transformed as:
\begin{equation}
\phi _{,xx}+3\phi _{,x}-\frac{1}{2\phi }\phi _{,x}^{2}+\frac{\dot{H}}{H^{2}}%
\phi _{,x}+\frac{2\kappa ^{2}}{3}\frac{\phi \left[ 2V\left( \phi
\right) -\left( 1+\phi \right) V^{\prime }\left( \phi \right)
\right] }{H^{2}}=0. \label{10}
\end{equation}
Now we  search for explicit solutions  that can be achieved by imposing the form of $V(\phi)$.

\subsubsection{Power-law accelerating models}

In order to arrive at accelerating models, i.e., $q=-\dot{H}/H^{2}-1<0$, we may assume that during the accelerating phase the deceleration parameter is a constant (in a pure de Sitter expansion the deceleration parameter is $-1$), so that
\begin{equation}\label{defq}
q=-\frac{\dot{H}}{H^{2}}-1=-q_{0}={\rm constant}\,.
\end{equation}
This expression can be simplified as
\begin{equation}
\frac{\dot{H}}{H^{2}}=q_{0}-1\,, \label{10a}
\end{equation}
which provides the form of the scale factor $a(t)$.
Furthermore, for simplicity, we impose on the potential the condition
\begin{equation}
2V\left( \phi \right) -\left( 1+\phi \right) V^{\prime }\left(
\phi \right) =0,
\end{equation}
that gives the solution
\begin{equation}
V(\phi )=V_{0}\left( 1+\phi \right) ^{2}.
\end{equation}

With these considerations, Eq.~(\ref{10}) reduces to
\begin{equation}
\phi _{,xx}+\left( 2+q_{0}\right) \phi _{,x}-\frac{1}{2\phi }\phi
_{,x}^{2}=0.  \label{11b}
\end{equation}

By denoting $d\phi /dx=u$, we obtain $d^{2}\phi
/dx^{2}=du/dx=\left( du/d\phi \right) \left( d\phi /dx\right)
=udu/d\phi $, so that Eq.~(\ref{11b}) becomes
\begin{equation}
u\frac{du}{d\phi }+\left( 2+q_{0}\right) u-\frac{1}{2\phi
}u^{2}=0,
\end{equation}
or equivalently
\begin{equation}
\frac{du}{d\phi }=\frac{1}{2\phi }u-\left( 2+q_{0}\right) ,
\end{equation}
which yields the following solution
\begin{equation}
u\left( \phi \right) =C_{1}\sqrt{\phi }-2\left( q_{0}+2\right)
\phi \,.
\end{equation}
Restoring the previous variables, one arrives at
\begin{equation}
\frac{d\phi }{C_{1}\sqrt{\phi }-2\left( q_{0}+2\right) \phi }=dx,
\end{equation}
\begin{equation}
x=\frac{1}{2\left( q_{0}+2\right) }\ln \frac{\phi }{\left( \sqrt{\phi }%
-2\left( q_{0}+2\right) \phi /C_{1}\right) ^{2}}-C_2.
\end{equation}

On the other hand, Eq.~(\ref{10a}) can be written as
\begin{equation}
\frac{d}{dt}\frac{1}{H}=1-q_{0},
\end{equation}
and then immediately integrated
\begin{equation}
H=\frac{\dot{a}}{a}=\frac{1}{\left( 1-q_{0}\right) t},
\end{equation}
to provide the solution
\begin{equation}
a(t)=a_0t^{1-q_{0}}.
\end{equation}
In this way we have obtained a power-law accelerating expansion depending on the value of $q_0$.

\subsection{The general case}

\subsubsection{Abel equation}

In the general case, Eq.~(\ref{10}) can be written as
\begin{eqnarray}\label{12a}
&&\phi _{,xx}+3\phi _{,x}-\frac{1}{2\phi }\phi _{,x}^{2}+\frac{1}{2}\frac{1}{%
1+\phi }\left\{ \frac{1}{\phi }\phi _{,x}^{2}+4\phi _{,x}+f(\phi
)\left( 1+\phi +\phi _{,x}+\phi _{,x}^{2}/4\phi \right) \right\}
\phi _{,x}+
 \nonumber\\
&&f(\phi )\left( 1+\phi +\phi _{,x}+\phi _{,x}^{2}/4\phi \right)
=0,
\end{eqnarray}
where we have denoted $U(\phi )=V^{\prime }/V$, and $f\left( \phi
\right) =2\phi \left[ 2-\left( 1+\phi \right) U(\phi )\right] $.
Eq. (\ref{12a}) can be written as
\begin{equation}
\phi _{,xx}+\frac{3}{2}(f+2)\phi _{,x}+\frac{3\phi \left( f+2\right) +f-2}{%
4\phi \left( 1+\phi \right) }\phi _{,x}^{2}+\frac{4+f}{8\phi
\left( 1+\phi \right) }\phi _{,x}^{3}+\left( 1+\phi \right) f=0.
\end{equation}
Introducing as above the variable $u$,
we obtain
\begin{equation}
u\frac{du}{d\phi }+\frac{3}{2}(f+2)u+\frac{3\phi \left( f+2\right) +f-2}{%
4\phi \left( 1+\phi \right) }u^{2}+\frac{4+f}{8\phi \left( 1+\phi \right) }%
u^{3}+\left( 1+\phi \right) f=0\,,
\end{equation}
and by dividing with $u^{3}$, we find
\begin{equation}
\frac{1}{u^{2}}\frac{du}{d\phi
}+\frac{3}{2}(f+2)\frac{1}{u^{2}}+\frac{3\phi
\left( f+2\right) +f-2}{4\phi \left( 1+\phi \right) }\frac{1}{u}+\frac{4+f}{%
8\phi \left( 1+\phi \right) }+\left( 1+\phi \right)
f\frac{1}{u^{3}}=0.
\end{equation}

By denoting $u=1/v$ we obtain an Abel equation of the form
\begin{equation}
\frac{dv}{d\phi }-\frac{3\phi \left( f+2\right) +f-2}{4\phi \left(
1+\phi
\right) }v-\frac{3}{2}(f+2)v^{2}-\left( 1+\phi \right) fv^{3}-\frac{4+f}{%
8\phi \left( 1+\phi \right) }=0\,.
\end{equation}
Our task is now to integrate such an equation.

\subsubsection{A simple accelerating solution}

The simplest case  corresponds to $f=0$, $V\left( \phi \right)
=V_{0}\left( 1+\phi \right) ^{2}$, and $\phi \ll 1$. Then the Abel
equation takes the form
\begin{equation}
\frac{dv}{d\phi }+\frac{1}{2\phi }v-3v^{2}-\frac{1}{2\phi }=0,
\end{equation}
with the general solution
\begin{equation}
v=\frac{\tan \left( \sqrt{6\phi }+C\right) }{\sqrt{6\phi }}.
\end{equation}

For small values of $\phi $, the solution can be approximated as
\begin{equation}
v=1+\frac{C}{\sqrt{6\phi }},\qquad u=\phi _{,x}=\frac{\sqrt{6\phi }}{\sqrt{6\phi }+C%
}.  \label{13}
\end{equation}

Since $\phi \ll 1$, one can approximate the potential as $V\approx
V_{0}={\rm constant}$. Then the condition for accelerating
expansion takes the form
\begin{equation}
\frac{\left[ 2+\phi _{,x}^{2}/\phi +4\phi _{,x}\right] \phi
_{,x}}{2\phi \left( 1+\phi _{,x}+\phi _{,x}^{2}/4\phi \right)
}+2\phi _{,x}>0,
\end{equation}%
which is obviously satisfied since $\phi _{,x}>0$ if $C>0$.

\subsection{Parameterizing by $q$}

In view of a cosmographic analysis of the models, it is convenient to parameterize the cosmological equations by the deceleration parameter $q$.  As discussed in \cite{ruth},
cosmography  is an extremely useful tool to discriminate among concurring cosmological models and, such a feature, is strictly related to  suitable parameterizations. In our case, the approach is very straightforward,   starting from Eqs.~(\ref{4a}) and (\ref{7a}).

Thus, the field equations given by Eqs.~(\ref{6b}) and (\ref{6c}), in terms of the deceleration parameter, can be recast in the following extremely simple form
\begin{eqnarray}
3H^{2}&=&\frac{\kappa ^{2}}{\left( 1+q\right) }\left[ 2V\left( \phi
\right) -\phi V^{\prime }\left( \phi \right) \right] ,  \label{8d}
 \\
\dot{H}&=&-\frac{\kappa ^{2}\left( 1-q\right) }{3\left( 1+q\right)
}\left[ 2V\left( \phi \right) -\phi V^{\prime }\left( \phi \right)
\right] , \label{8e}
\end{eqnarray}
where the deceleration parameter is defined according to the sign convention of Eq.~(\ref{defq}).
Therefore, the cosmological field equations can be expressed in
terms of the scalar field, the scalar field potential, and
the deceleration parameter. However, the Klein-Gordon equation, given by Eq.~(\ref{3a}), also needs to be taken into account. Let us work out some simple examples adopting this parameterization.

\subsubsection{Exponential expansion}

In the case of the exponential expansion $q=-1$, $H=H_{0}$ = constant, $\dot{H%
}=0$, and $a(t)=a_{0}\exp \left[ H_{0}\left( t-t_{0}\right)
\right] $,
respectively. For this case, Eq.~(\ref{8e}) is automatically satisfied, while Eq.~(\ref{8d}%
) becomes
\begin{equation}
\phi V^{\prime }\left( \phi \right) -2V\left( \phi \right)
+\frac{6}{\kappa ^{2}}H_{0}^{2}=0,
\end{equation}
which provides the following solution
\begin{equation}
V\left( \phi \right) =\frac{3}{\kappa ^{2}}H_{0}^{2}+\frac{V_{0}}{\kappa ^{2}%
}\phi ^{2},
\end{equation}%
where $V_{0}$ is an arbitrary constant of integration.
The general solution for the scalar field is obtained from the Klein-Gordon equation, which is given by
\begin{equation}
\phi \left( t\right) =\left[ \frac{3H_{0}}{\exp \left[ H_{0}\left(
t-t_0\right) \right] \pm \sqrt{3V_{0}}}\right] ^{2}.
\end{equation}
The physical solution corresponds to the choice of the positive sign,
in order to avoid any singular behavior of the scalar field.

\subsubsection{Power law accelerated expansion}

As above, in the case of  power law accelerated expansion, we have the constant $q=q_{0}$ and
\begin{equation}
H(t)=\frac{1}{\left( 1-q_{0}\right) t},\qquad \dot{H}=-\frac{1}{1-q_{0}}\frac{1}{%
t^{2}},\qquad \mbox{with }\quad q_{0}\neq 1.
\end{equation}
The field equations, given by Eqs.~(\ref{8d}) and (\ref{8e}), provide the following expression
\begin{equation}
\frac{1}{(1-q_{0})t^2}=\frac{\kappa ^{2}\left( 1-q_{0}\right) }{%
3\left( 1+q_{0}\right) }\left[ 2V\left( \phi \right) -\phi
V^{\prime }\left( \phi \right) \right] .
\end{equation}
Hence, the field equations reduce to a single condition for the
potential, given by
\begin{equation}
2V\left( \phi \right) -\phi V^{\prime }\left( \phi \right)
=\frac{3\left( 1+q_{0}\right) }{\left( 1-q_{0}\right)
^{2}}\frac{1}{t^{2}}.  \label{11}
\end{equation}
Reversing the argument, any form of scalar field potential (and then the $f(X)$-function) compatible with power law expansion ${\displaystyle a(t)=a_0 t^{1-q_0}}$ can be achieved in this way.

%

\section{Cosmological perturbations}
\label{sec:pert}

\subsection{Field equations and conservation laws}

To understand the implications of these models in the context of structure formation, we will derive the perturbation equations and analyze them in some
specific cases of interest. This paves the way for a detailed comparison of the predictions with the cosmological data on large scale structure and the cosmic microwave background. For generality, we will keep the parameter $\Omega_A$ in the formulas in this section.

We work in the Newtonian gauge \cite{Ma:1995ey}, which can be parameterized by the two gravitational potentials $\Phi$ and $\Psi$,
\be
ds^2= -\lp 1+2\Psi \rp dt^2+a^2(t)\lp 1+2\Phi\rp d\vec{x}^2\,. 
\ee
The $0-0$ part of the field equations is
\ba
\frac{k^2}{a^2}\Phi 
+ 3\lp H-\frac{\dot{\phi}}{2\Fp}\rp\dot{\Phi} 
- 3\lp H^2 +\frac{H\dot{\phi}}{\F}-\frac{\dot{\phi}^2}{4\phi\Fp}\rp\Psi   
   \nonumber \\ 
=\frac{1}{2\F}\lb \kappa^2\delta\rho_m + \lp \frac{3}{4\phi^2}\dot{\phi}^2+V'(\phi)-3H^2-\frac{k^2}{a^2}\rp\varphi - 3\lp H + \frac{\dot{\phi}}{2\phi}\rp 
\dot{\varphi}\rb\,,
\ea
where we have denoted $\varphi=\delta\phi$.
The Raychaudhuri equation for the perturbations reads
\ba
\lb 6\lp H^2+2\dot{H}\rp - 2\frac{k^2}{a^2} + \frac{6}{\F}\lp 
\ddot{\phi}-\frac{\dot{\phi}^2}{\phi^2}+H\dot{\phi}\rp \rb \Psi 
+ 3\lp 2H - \frac{\dot{\phi}}{\F}\rp \lp \dot{\Phi} - \dot{\Psi}\rp 
-6\ddot{\Phi} 
 \nonumber \\
= \frac{1}{\F}\lb \kappa^2\lp \delta\rho_m+3\delta p_m\rp 
+\lp 6H^2+6\dot{H} + 3\frac{\ddot{\phi}}{\phi^2}-2V'(\phi)+\frac{k^2}{a^2}\rp\varphi
+3\lp H-\frac{2\dot{\phi}}{\phi}\rp \dot{\varphi}
+ 3\ddot{\varphi}
\rb\,.
\ea
The $0-i$ equation is
\be
- \lp H+\frac{\dot{\phi}}{2\Fp}\rp\Phi + \dot{\Phi}
= \frac{1}{2\Fp}\lb \kappa^2\lp\rho_m+p_m\rp a v_m + \lp H+\frac{3\dot{\phi}}{2\phi}\rp\varphi+\dot{\varphi}\rb\,.
\ee
Note that the set of perturbed field equations is completed by the off-diagonal spatial piece:
\be
\Psi+\Phi=-\frac{\varphi}{\F}\,.
\ee
Assuming a perfect fluid, the continuity and Euler equations for the matter component are 
\ba
\dot{\delta} + 3H\lp c_s^2-w\rp\delta & = & -\lp 1+w \rp \lp 3\dot{\Phi}-\frac{k^2}{a}v\rp\,, \\
\ddot{v}+\lp 1-3c_a^2\rp Hv & = &  \frac{1}{a}\lp \Psi + \frac{c_s^2}{1+w}\delta\rp\,, 
\ea
respectively. The linear part of the Klein-Gordon equation is then compatible with the above system. For completeness, it is 
\ba
\ddot{\varphi}+\lp 3H+\frac{1}{\phi}\rp \dot{\varphi} + \lp \frac{k^2}{a^2}+\frac{\dot{\phi}^2}{2\phi^2}- \frac{2}{3}V''(\phi)\rp\varphi &= & \nonumber
\lp 2\ddot{\phi}+6H\dot{\phi}-\frac{3}{2\phi}\dot{\phi}^2\rp\Psi + \dot{\phi}\lp \dot{\Psi}-3\dot{\Phi}\rp - \frac{\phi}{3}\delta R\,.
\ea
This completes the presentation of the field equations and the conservation laws.

\subsection{Matter dominated cosmology}

Let us consider formation of structure in the matter-dominated universe, where $w_m=c_s^2=0$. 
In this subsection, we shall consider scales deep inside the Hubble radius. This approximation is well known in the literature and indeed we will arrive at a similar result as in e.g. \cite{Boisseau:2000pr}.
Note that then the spatial gradients are more important than the time derivatives and, consequently, the 
matter density perturbations are much stronger than the gravitational potentials. Combining the continuity and the Euler equation at this limit, one obtains
\be
\ddot{\delta}=-2H\dot{\delta}-\frac{k^2}{a^2}\Psi\,.
\ee  
We need then to solve the gravitational potential. Let us define $\Pi = a^2\rho_m\delta_m/k^2$ and write the field equations and the Klein-Gordon equation at this limit in a very simple way as
\ba
\Fp\Phi &=& \Pi-\varphi\,, \\
\Fp\lp\Psi+\Phi\rp & = & -\varphi\,, \\
-2\Fp\Psi & = & \Pi + \varphi\,, \\
3\varphi & = & -2\phi\lp \Psi + 2\Phi\rp\,.
\ea
We immediately see that one of the equations is (as expected) redundant, and that the potential $\Psi$ is (as usual) proportional to $\Pi$, where now the proportionality is given as a function 
of the field $\phi$. Our result is
\be \label{delta_evol}
\ddot{\delta} + 2H\dot{\delta} = 4\pi G_{\rm eff}\rho_m\delta\,,
\ee
with
\be \label{delta_evol2}
 G_{\rm eff} \equiv \frac{\OA-\frac{1}{3}\phi}{\OA\Fp}G\,. 
\ee
This shows that there are no instabilities in the evolution of the matter inhomogeneities, in contrast to the Palatini-$f(\R)$ models and some matter-coupled scalar field models (recall our theory 
can be mapped into such in the Einstein frame).

\subsection{Vacuum fluctuations}

The propagation of our scalar degree of freedom in vacuum is also a crucial consistency check on the theory. Let us set $\rho_m=0$. Let us consider the curvature perturbation in the uniform-field 
gauge $\zeta$. In terms of the Newtonian gauge perturbations this is 
\be
\zeta= \Phi-\frac{H}{\dot{\phi}}\varphi\,.
\ee
After considerably more tedious algebra than in the previous case, we obtain the exact (linear) evolution equation
\be \label{eta_evol}
\ddot{\zeta}+\lb 3H
-2\frac{\ddot{\phi}+2\dot{H}\Fp -\frac{\dot{\phi}^2}{\F}}{\dot{\phi}+2H\Fp} 
+ \frac{\phi\Fp}{\dot{\phi}^2}\lp\frac{2\ddot{\phi}\dot{\phi}}{\phi\Fp}
 + \frac{\dot{\phi}^3\Fp^2\phi}{1-\phi^3\Fp^3}\rp\rb\dot{\zeta} = - \frac{k^2}{a^2}\zeta\,.
\ee
The friction term depends on the perturbation variable we consider, but the sound speed term retains its 
form. Thus perturbations at small scales propagate with the speed of light, as in canonic scalar field 
theory. This excludes also instabilities in the scalar field perturbations. Now equation (\ref{eta_evol}) can be used to study 
generation of fluctuations in $f(X)$-inflation. Construction of specific models and their observational tests are left for forthcoming studies.

\section{Conclusions}\label{sect5}

In this paper, we have discussed cosmological applications of hybrid metric-Palatini $f(X)$-gravity. This theory differs from Palatini $f(\R)$-gravity for two main aspects: $i)$  both metric and Palatini approach are considered including $R$ and $f(\R)$ in the gravitational action; $ii)$ deviations from standard general relativity are considered including the field $X=\kappa^2 T+ R$ where $X$ is exactly zero in the Einstein case. This approach is particularly relevant in view of the so-called {\it chameleon mechanism} useful to parameterize the astrophysical scales according to the cosmic densities: At short scales (Solar System scales) $X\rightarrow 0$, while it increases at larger scales since dark matter and dark energy effects have to be included. Another option is to consider the field dynamically evolving to $X\rightarrow 0$ today, which renders
cosmological dynamics compatible without resorting to the chameleon mechanism.
Considering $f(X)$-gravity means  not distinguishing {\it a priori} matter and geometric sources, but taking into account both of them into the dynamics under the same standard. Furthermore, shortcomings connected to the conformal transformations are avoided. For example, the singular cases $w=0$ for metric $f(R)$-gravity and $w=-3/2$ for Palatini $f(\R)$-gravity in the scalar-tensor representation of these theories are avoided, $f(X)$ gravity being naturally metric-affine. This means that our hybrid model can be fully recast into a scalar-tensor theory of gravity, where the kinetic and potential terms are well defined. Furthermore, it is straightforward to recover general relativity, which is fully restored as soon as $\phi\rightarrow 0$ (see Eq.~(\ref{eq:S_scalar2})).

Here, we have also discussed the FRW cosmology coming out from $f(X)$-gravity. Accelerating expansion, and, in general, any cosmological behavior, strictly depends on the effective scalar field potential, which assigns the form of $f(X)$.  In particular, we have examined toy models where the deceleration parameter evolves according to such a potential.  This approach is particularly useful in view of a cosmographic analysis  to discriminate realistic models by observations (see \cite{cosmography} for $f(R)$-gravity in metric approach). However, we conclude in noting that this is only a seminal paper. An extensive study of $f(X)$-cosmology and its comparison with observations, amongst other issues, will be the topic of forthcoming papers.

\section*{Acknowledgments}
SC is supported by INFN (iniziativa specifica NA12). TSK is supported by the Research Council of Norway. The work of TH is
supported by an RGC grant of the government of the Hong Kong SAR. FSNL acknowledges financial support of the Funda\c{c}\~{a}o para a Ci\^{e}ncia e Tecnologia through the grants CERN/FP/123615/2011 and CERN/FP/123618/2011. GJO is supported by the Spanish grants FIS2008-06078-C03-02, FIS2011-29813-C02-02, the Consolider Programme CPAN (CSD2007-00042), and the JAE-doc program of the Spanish Research Council (CSIC).


\begin{thebibliography}{99}

\bibitem{Harko:2011nh}
  T.~Harko, T.~S.~Koivisto, F.~S.~N.~Lobo and G.~J.~Olmo,  Phys.\ Rev.\ D {\bf 85}, 084016 (2012).
 \bibitem{eps}
 S. Capozziello, M. De Laurentis, L. Fatibene, M. Francaviglia, Int. J. Geom. Meth. Mod. Phys. {\bf 9},  1250072 (2012).

 \bibitem{camel}
 J. Khoury and A. Weltman, Phys. Rev. Lett.  {\bf 93}, 171104
(2004);  S. Capozziello and S. Tsujikawa, Phys.Rev. {\bf D 77},  107501 (2008).

\bibitem{Koivisto:2012za}
  T.~S.~Koivisto, D.~F.~Mota and M.~Zumalacarregui,
  arXiv:1205.3167 [astro-ph.CO].


\bibitem{expansion}
S.~Perlmutter {\it et al.}, Astrophys.\ J.\  {\bf 517}, 565
(1999); A.~G.~Riess {\it et al.}, Astron.\ J.\  {\bf 116}, 1009
(1998); A.~G.~Riess {\it et al.}, Astrophys.\ J.\  {\bf 607}, 665
(2004); A. Grant {\it et al}, Astrophys. J. {\bf 560} 49-71
(2001); S. Perlmutter, M. S. Turner and M. White, Phys. Rev. Lett.
{\bf 83} 670-673 (1999); C. L. Bennett {\it et al}, Astrophys. J.
Suppl. {\bf 148} 1 (2003);
G. Hinshaw {\it et al}, [arXiv:astro-ph/0302217];
E.~J.~Copeland, M.~Sami and S.~Tsujikawa,
  Int.\ J.\ Mod.\ Phys.\  D {\bf 15}, 1753 (2006).
 P. J. E. Peebles and B. Ratra, Rev. Mod. Phys. \textbf{75}, 559 (2003); T. Padmanabhan, Phys. Repts. \textbf{380}, 235 (2003);
  G. Hinshaw et al.,  Astrophys. J. Supplement {\bf 180},  225 (2009).

\bibitem{fRgravity} S. Capozziello, Int. J. Mod. Phys. {\bf D 11}, 483  (2002);
S. Nojiri and S.D. Odintsov, Int.J.Geom.Meth.Mod.Phys. {\bf 4}, 115  (2007);
S. Capozziello and M. Francaviglia, Gen. Rel. Grav. {\bf 40}, 357  (2008);
 T.~P.~Sotiriou and V.~Faraoni, Rev. Mod. Phys. {\bf 82}, 451 (2010);
A.~De Felice and S.~Tsujikawa, Living Rev.\ Rel.\  {\bf 13}, 3 (2010);
F.~S.~N.~Lobo, arXiv:0807.1640 [gr-qc].  

\bibitem{revnoi}
S. Capozziello and M. De Laurentis, Phys. Rept.  {\bf 509}, 167  (2011).

\bibitem{cqg}
S. Capozziello, R. de Ritis, A. A. Marino,  Class. Quantum Grav. {\bf 14},   3243 (1997).

\bibitem{libro}
S. Capozziello and V. Faraoni, {\it Beyond Einstein gravity: A Survey of gravitational theories for cosmology and astrophysics}, Fundamental Theories of Physics, Vol. 170, Springer, 2010, New York.

\bibitem{Amendola:2010bk}
  L.~Amendola, K.~Enqvist, T.~Koivisto,
  Phys.\ Rev.\  {\bf D83}, 044016 (2011);
  T.~S.~Koivisto,
  Phys.\ Rev.\ D {\bf 83} (2011) 101501;
 T.~S.~Koivisto,
  Phys.\ Rev.\ D {\bf 84} (2011) 121502.

\bibitem{Olmoetal}
G.~J.~Olmo,
  JCAP {\bf 1110}, 018 (2011);
 C.~Barragan and G.~J.~Olmo,
  Phys.\ Rev.\ D {\bf 82}, 084015 (2010);
 G.~J.~Olmo, H.~Sanchis-Alepuz and S.~Tripathi,
  Phys.\ Rev.\ D {\bf 80}, 024013 (2009).

\bibitem{allemandi}
G.~Allemandi, M.~Capone, S.~Capozziello, and M.~Francaviglia,
Gen.\ Rel.\ Grav.\  {\bf 38}, 33 (2006).

\bibitem{dark}
S.~Capozziello, M. De Laurentis, M.~Francaviglia and S.~Mercadante,
Found.\ Phys.\  {\bf 39}, 1161 (2009)

\bibitem{Flanagan:2003iw}
  E.~E.~Flanagan,
  Class.\ Quant.\ Grav.\  {\bf 21}, 417 (2003).


\bibitem{Olmo:2011uz}
  G.~J.~Olmo,
  [arXiv:1101.3864 [gr-qc]].

\bibitem{Biswas:2011ar}
  T.~Biswas, E.~Gerwick, T.~Koivisto and A.~Mazumdar,
  Phys.\ Rev.\ Lett.\  {\bf 108} (2012) 031101.


\bibitem{Harko:2010mv}
 T.~Harko and F.~S.~N.~Lobo,
  Eur.\ Phys.\ J.\  C {\bf 70}, 373 (2010);
%
  T.~Koivisto,
  Class.\ Quant.\ Grav.\  {\bf 23}, 4289-4296 (2006).
  [gr-qc/0505128].
%
  O.~Bertolami, C.~G.~Boehmer, T.~Harko and F.~S.~N.~Lobo,
  Phys.\ Rev.\  D {\bf 75}, 104016 (2007);
%
O. Bertolami and J. P\'aramos, Phys.\ Rev.\  D {\bf 77}, 084018
(2008);
%
  O.~Bertolami, F.~S.~N.~Lobo and J.~Paramos,
  Phys.\ Rev.\  D {\bf 78}, 064036 (2008);
%
O.~Bertolami, J.~Paramos, T.~Harko and F.~S.~N.~Lobo,
  arXiv:0811.2876 [gr-qc].
%
 T.~Harko, T.~S.~Koivisto and F.~S.~N.~Lobo,
  Mod.\ Phys.\ Lett.\ A {\bf 26}, 1467 (2011). 


\bibitem{Poplawski:2006ey}
  N.~J.~Poplawski,
  arXiv:gr-qc/0608031;
  T.~Harko, F.~S.~N.~Lobo, S.~'i.~Nojiri and S.~D.~Odintsov,
  Phys.\ Rev.\ D {\bf 84}, 024020 (2011).  



\bibitem{Koivisto:2005yc}
  T.~Koivisto and H.~Kurki-Suonio,
  Class.\ Quant.\ Grav.\  {\bf 23}, 2355 (2006);
  T.~Koivisto,
  Phys.\ Rev.\ D {\bf 73}, 083517 (2006).


\bibitem{Olmo:2006zu} 
 G.~J.~Olmo, Phys.\ Rev.\ D {\bf 77}, 084021 (2008);
  Phys.\ Rev.\ Lett.\  {\bf 98}, 061101 (2007).

\bibitem{Olmo:2005jd}
  G.~J.~Olmo,
  Phys.\ Rev.\  {\bf D72}, 083505 (2005); Phys.\ Rev.\ Lett.\  {\bf 95}, 261102 (2005).  

\bibitem{Koivisto:2009jn}
  T.~S.~Koivisto,
  AIP Conf.\ Proc.\  {\bf 1206} (2010) 79.

\bibitem{Boisseau:2000pr}
  B.~Boisseau, G.~Esposito-Farese, D.~Polarski and A.~A.~Starobinsky,
  Phys.\ Rev.\ Lett.\  {\bf 85} (2000) 2236
  [gr-qc/0001066].

\bibitem{stt} 
  P.~Brax, C.~van de Bruck, D.~F.~Mota, N.~J.~Nunes and H.~A.~Winther,
  Phys.\ Rev.\ D {\bf 82}, 083503 (2010)
  [arXiv:1006.2796 [astro-ph.CO]].
  P.~Brax, A.~-C.~Davis, B.~Li and H.~A.~Winther,
  Phys.\ Rev.\ D {\bf 86}, 044015 (2012)
  [arXiv:1203.4812 [astro-ph.CO]].



\bibitem{ruth}
  S. Capozziello, R. Lazkoz,  V. Salzano,
  Phys.\ Rev.\  {\bf D84}, 124061 (2011).

\bibitem{Ma:1995ey}
  C.~-P.~Ma and E.~Bertschinger,
  Astrophys.\ J.\  {\bf 455} (1995) 7.


\bibitem{cosmography}
S. Capozziello, V.F. Cardone, V. Salzano,
Phys.  Rev.  {\bf D78},  063504  (2008).



\end{thebibliography}
\end{document}